# Meteoritic Tutton salt, a naturally inspired reservoir of cometary and asteroidal ammonium

Sergey N. Britvin[1,2]*, Oleg S. Vereshchagin[1], Natalia S. Vlasenko[1], Maria G. Krzhizhanovskaya[1] & Marina A. Ivanova[3]

[1]Saint-Petersburg State University, Universitetskaya Nab. 7/9, 199034 St. Petersburg, Russia
[2]Kola Science Center, Russian Academy of Sciences, Fersman Str. 14, 184200 Apatity, Russia
[3]Vernadsky Institute of Geochemistry of the Russian Academy of Sciences, Kosygin St. 19, Moscow 119991, Russia

* Corresponding author. E-mail: sergei.britvin@spbu.ru

The lack of benchmark data on the real minerals – native ammonium carriers in Solar System gives rise to controversial opinions on extraterrestrial ammonium reservoirs. We herein report on discovery of the first mineral carrier of meteoritic ammonium and show its relevance to the compositional and spectral characteristics of cometary and asteroidal bodies. Chemically distant from previously inferred volatile organics or ammoniated phyllosilicates, it is an aqueous metal-ammonium sulfate related to a family of so-called Tutton salts. Nickeloan boussingaultite, $(NH_4)_2(Mg,Ni)(SO_4)_2·6H_2O$, occurs in Orgueil, a primitive carbonaceous chondrite closely related to (162173) Ryugu and (101955) Bennu, the C-type asteroids. The available spectroscopic, chemical and mineralogical data signify that natural Tutton salts perfectly fit into the role of ammonium reservoir under conditions of cometary nuclei and carbonaceous asteroids.

Elucidating the evolution of cosmic nitrogen is a persistent challenge of modern cosmochemistry[1]. Ammonia and ammonium, the simplest carriers of reduced, life-essential nitrogen, are believed to be abundant on the icy Solar System bodies[2-5]. Spectroscopic observations indicate the likely occurrence of ammonium salts in surficial deposits of (1) Ceres, a dwarf planet[6,7], and in cometary bodies[5,8]. However, interpretations based on remote sensing and laboratory simulations[5,7,9,10] remain ambiguous without the knowledge of real minerals – natural ammonium carriers that might constitute the celestial bodies.

The absence of recognizable ammonium minerals in meteorites, cometary and asteroid sample returns lead to the opinion that such indigenous salts could disappear – either sublimated or thermally decomposed already in space[5,11]. The questionable point of such opinion is that a variety of aqueous minerals, many of them as unstable as ammonium ones, are successfully delivered on Earth with meteorite influx[12]. In addition, traces of ammonium were confirmed in primitive meteorites – aqueously altered carbonaceous chondrites[2,3] and in the samples returned from Ryugu, a C-type asteroid[13,14]. Unfortunately, all ammonium determinations in the samples of extraterrestrial origin are based on the analyses of leachates or evolved gases released from bulk probes. The identification of real ammonium minerals is an uncharted area of space research. It may take place that native ammonium salts, being much scarce than aqueous minerals, are simply overlooked during conventional analytical procedures. A recent report of ammonium mineral from Lunar regolith[15] supports such an assumption. To elucidate the problem, our team re-examined the material of Orgueil, a carbonaceous chondrite that contains ammonium of definitely extraterrestrial origin[16]. We herein report the discovery of the first meteoritic ammonium mineral and demonstrate that it may serve as the most likely carrier of bound ammonia in cometary and asteroidal bodies.

Orgueil, a witnessed fall (1864, Midi-Pyrenees, France), is a biggest (13 kg) carbonaceous chondrite belonging to most primitive meteorites incorporated into the C1I (Ivuna) group[17]. The latter, consisting just of 9 meteorites, attracts enormous interest owing to mineralogical and chemical features, such as extremely high water content[18] and occurrence of diverse organic compounds of extraterrestrial origin, including nitrogen-bearing moieties[19,20]. Orgueil is the closest relative of the samples recently returned from asteroids (162173) Ryugu and (101955) Bennu[21-23]. Due to its large mass, Orgueil is the most accessible C1I chondrite distributed among museums and laboratories worldwide[24]. Owing to the striking compositional similarity to the solar photosphere, Orgueil is an international standard used for normalization of elemental abundances in geochemical and cosmochemical studies[25]. The meteorite, like other CI1 chondrites, consists entirely of matrix, with no traces of chondrules or Ca,Al-rich inclusions (CAIs). The matrix constituents are Mg-rich, Fe-bearing serpentine and smectite, and a poorly crystallized ferrihydrite (a ferric hydroxide)[26]. Accessory minerals are comprised by irregular grains of Ca-Mg-Fe carbonates, olivine, pyroxene, sulfides. Sulfate minerals, in form of veinlets infilling the cracks in the matrix, are represented by gypsum, magnesium sulfate hydrates, sometimes with admixture of Na and Ni[27].

Ammonium-bearing salts (probably sal ammoniac) were originally described on the fusion crust of Orgueil immediately after the fall[28-30], but were not revealed by subsequent studies, leading to the assumption that they were lost upon meteorite storage[17]. Our results indicate that the source of surficial ammonium evaporates detected by Daubrée (1864) and Cloëz (1864)[28,29], as well as of water-soluble ammonium in Orgueil leachates[16] is nickeloan boussingaultite, $(NH_4)_2(Mg,Ni)(SO_4)_2 \cdot 6H_2O$. We identified this mineral as an accessory phase in the studied Orgueil sample. The mineral occurs as euhedral isometric, isolated faceted crystals and their intergrowths up to 30 μm in size (Fig. 1) disseminated within phyllosilicate matrix. Of the 4-5 mm meteorite piece crushed between the glass slides, one could recognize eight Ni-boussingaultite crystals. The mineral is colorless with a distinct greenish tint due to significant amount of Ni in its composition. The primary phase identification was carried out by means of X-ray single-crystal study. The best of handpicked crystals (Fig. 1a) had the following crystallographic parameters: monoclinic, space group $P2_1/c$ (# 14), $a$ 6.2137(3), $b$ 12.5213(7), $c$ 9.2526(5) Å, $\beta$ 106.934(6)°, $V$ 688.67(7) Å$^3$, $Z$ = 2. The obtained data evidenced that the mineral belongs to hydrated sulfates of the general formula $A^+_2B^{2+}(SO_4)_2 \cdot 6H_2O$, where $A^+$ is either ammonium or alkali cation (except for Na$^+$ and Li$^+$); $B^{2+}$ is a medium-sized metal cation. In mineralogy, these sulfates are incorporated into the picromerite group, whereas in chemistry and materials science they are known as the Tutton's or Tutton salts[31-33].

Subsequent crystal structure solution and refinement showed that cation $A^+$ in the mineral from Orgueil is comprised by NH$_4^+$, whereas the $B$-site has a mixed occupancy intermediate between Mg$^{2+}$ and either Fe$^{2+}$ or Ni$^{2+}$. Electron microprobe data obtained from the same crystal evidenced that the mineral does not contain Fe but bears a significant amount of Ni substituting for Mg (Fig. 1b). This allowed finalization of the structure refinement, which converged with the chemical formula $(NH_4)_2(Mg_{0.65}Ni_{0.35})_{1.00}(SO_4)_2 \cdot 6H_2O$. The refined Mg/Ni atomic ratio, 0.65/0.35, well agrees with the value of 0.63/0.37 obtained from electron microprobe data. Crystal structure of the studied crystal is illustrated in Fig. 2. Infrared spectroscopy confirms all features of NH$_4$-Mg and NH$_4$-Ni Tutton salts[34], in particular the presence of absorption bands characteristic of ammonium ion: 2888, 2840 cm$^{-1}$ (stretching modes of N–H bonds) and 1471, 1432 cm$^{-1}$ (bending vibrations of NH$_4^+$) (Fig. 3a, Table 1).

The discovery of the first meteoritic ammonium mineral confirms the existence of suspected but previously unidentified crystalline ammonium salts in Orgueil. Cloëz (1864)[29] reported 0.1 wt.% of ammonium in aqueous leachates collected shortly after the witnessed fall. The latest analytical data almost exactly replicate the above value, giving a mean of 0.068 ± 0.015 wt% NH$_4$ in leachates[16]. In view that Orgueil, likewise other CI1 chondrites, experienced terrestrial alteration[35,36], it is significant that this water-soluble ammonium has distinct extraterrestrial origin, with δ$^{15}$N ranging from +37 to +132 ‰ [ref. 16], consistent with δ$^{15}$N values in very fresh sample returns from asteroids Bennu (+57 to 82 ‰) and Ryugu (+43 ‰)[23,37]. Therefore, notwithstanding for the possible terrestrial redistribution of sulfates[35], nickeloan boussingaultite represents the indigenous ammonium mineral in Orgueil.

In view of recent advances in space research, in particular the challenge of ammonium reservoir in comets and asteroids, we attempted a comparison of infrared spectra of comet 67P/Churyumov-Gerasimenko[38] with corresponding spectra of nickeloan boussingaultite from Orgueil. The choice of comet 67P is connected with the availability of a quality spectrum in the 2.5 – 4 μm region[39] that supposedly evidences for ammonium salts in the composition of 67P[5,40].

In the context of the problem, it is important that infrared spectra of ammonium Tutton salts, $(NH_4)_2B^{2+}(SO_4)_2 \cdot 6H_2O$, exhibit significant red shift of the bands corresponding to O–H stretching vibrations relative to ammonium-free counterparts[34,41]. The effect is caused by incorporation of ammonium, which becomes involved in a pronounced network of cross-linked hydrogen bonds (Fig. 2). The O–H bands offset towards smaller wavenumbers (i.e., longer wavelengths) reaches 150–200 cm$^{-1}$. As a result, the O–H stretching bands in the IR-spectrum of nickeloan boussingaultite from Orgueil match the position of a so-called "3.2 μm feature" observed in the infrared spectra of some comets and asteroids[5]. This enigmatic split band[42] was assigned to ammonium in cometary nucleus of 67P[5], because its position and shape corresponds to N–H stretching vibrations in the spectra of simulants prepared from the mixtures of pyrrhotite, $Fe_{1-x}S$, and anhydrous ammonium salts[5]. Several candidate compounds were checked for the role of ammonium reservoir in comet 67P, the best suited one was ammonium formate, $HCOONH_4$ [ref. 5]. Our infrared reflectance data (Fig. 3b, Table 2) evidence that the 3.2 μm feature can be equally explained by the presence of $NH_4$-bearing Tutton salts. Curiously in this case, the 3.2 μm band is not assigned to ammonium, but to crystal hydrate water. However, its red shift is induced by, and therefore evidences for the presence of ammonium.

Cometary mineralogy share all basic features with primitive carbonaceous chondrites and their closest relatives – carbonaceous asteroids (162173) Ryugu and (101955) Bennu[21-23,43-45]. In the context of the present work, it is notably that all these bodies contain noticeable amounts of sulfides, including pyrrhotite, $Fe_{1-x}S$, pentlandite, $(Fe,Ni)_9S_8$, and cubanite, $CuFe_2S_3$. The former two sulfides bear notable to very high Ni contents[46]; therefore, they might serve as a source of Ni for Tutton salts. Nickelboussingaultite, $(NH_4)_2(Ni,Mg)(SO_4)_2 \cdot 6H_2O$ where Ni>Mg, is known on Earth as an aqueous alteration product of Ni-sulfide ores[47,48]. The origin of nickel-rich boussingaultite in Orgueil could be invoked by analogous alteration of Ni-bearing sulfides. This corroborates with the known pronounced aqueous alteration of carbonaceous asteroids[21-23]. The presence of abundant molecular oxygen and $SO_2$ in the composition of the coma of 67P[49] supports the likelihood of the low-temperature aqueous oxidation of sulfides. Phyllosilicates would act as a source of Mg, whereas Fe, being predominant in sulfides, is dumped out into ferric iron-bearing minerals abundant both in Orgueil and asteroid Ryugu[21,22,26,50]. This mechanism explains the emergence of Mg-Ni sulfates completely devoid of Fe. In this respect, the latest finding of iron-free newberyite, $MgHPO_4 \cdot 3H_2O$, in the fresh samples returned from asteroid Bennu[23] evidences that oxidative Mg/Fe aqueous fractionation also took place on this asteroid. The only but significant feature that can not be explained by the in-situ oxidation of native sulfides is the complete absence of Cu in sulfates of Orgueil. Cubanite $CuFe_2S_3$, an accessory phase in both Orgueil and comet Wild2 [ref. 45,51,52] occurs in Orgueil as very fresh, sharp, unaltered twinned crystals (Fig. 6). It appears unlikely that cubanite was not affected by oxidative alteration whereas pentlandite, being quite resistant to oxidation, would oxidize, acting as a source of Ni released into sulfates. Therefore it is also possible that nickeloan boussingaultite in Orgueil is in fact a remnant of the early solutions which were not directly connected with the oxidation of Fe-Ni sulfides.

The 3.2–3.1 μm feature similar to that found in the infrared spectrum of comet 67P was detected in the spectra of (1) Ceres and several asteroids[5-7]. It was suggested that, together with a strong 2.7 μm band, this argues for the wide distribution of ammoniated phyllosilicates in these space bodies[7]. We suggest that like in case of comet 67P, the infrared spectra of (1) Ceres can be interpreted in terms of superposition of ordinary (ammonium-free) phyllosilicates abundant in carbonaceous chondrites and asteroids[7,21-23], and ammonian Tutton salts like that discovered in our work, with no necessity of implementation of exotic ammoniated clays.

Discovery of the first ammonium mineral in meteoritic matter might call for the reassessment of the analytical techniques implemented to extraterrestrial objects. It is shown in our paper that crystalline ammonium minerals can be scarce in meteoritic, cometary or asteroidal matter. They are generally water-soluble. In addition, these phases may have very low thermal decomposition thresholds. According to our observations, the irreversible loss of crystal hydrate water from ammonian Tutton salts proceeds in the range of 80 – 100 °C. The rarity and inherent instability of ammonium minerals under conditions of focused electron/ion beam in vacuum (electron or ion microprobe) may result in the omissions of these important phases during regular analytical procedures.

The existence of Tutton salts in both cometary nuclei and carbonaceous asteroids looks quite arguable, based on the fact of their natural occurrence in mineralogically relevant carbonaceous chondrite. From this point of view, the naturally confirmed mineral appears to be a potent candidate for the role of ammonium reservoir, in comparison with other proposed ephemeral compounds not yet confirmed in nature.

**Methods**
**Samples**
The 4×4×5 mm$^3$ piece of Orgueil from the collections of Vernadsky Institute of Geochemistry was crushed between flat glass plates. The material was inspected under binocular microscope at 120× magnification. The boussingaultite crystals could be distinguished from other near-colorless minerals (dolomite, hexahydrite, gypsum, sulfur) due to euhedral isometric crystal shape and the greenish tint. The crystals were handpicked from the sample for subsequent study.

**X-ray single-crystal study and crystal structure**

The single-crystal reflection dataset was collected by means of a Rigaku Oxford Diffraction XtaLAB Synergy-S diffractometer equipped with a microfocus X-ray tube (Mo$K\alpha$) and HyPix-6000 hybrid photon counting detector. Data collection, processing and reduction routines were performed with CrysAlisPro v.171.42 software (Rigaku Oxford Diffraction 2023). The crystal structure was solved and refined to $R_1$ = 0.029 using *SHELX*-2018 set of programs[53] incorporated into Olex2 v.1.5 graphical user interface[54]. The results of crystallographic study are provided in Supplementary Tables S1–S5.

**Electron microprobe analysis**

The mineral was found to be extremely unstable under the electron beam. In particular, application of wavelength-dispersive analyzer (WDX) was not possible, owing to instantaneous decomposition (burning out) of the crystals even under the de-focused beam. The energy-dispersive analysis (EDX) was carried out on the same crystal (Fig. 1a) which was used for X-ray single crystal study, in order to cross-check the Mg/Ni ratios determined by these two independent methods. The analyzed crystal was glued onto carbon tape and vacuum-coated by conductive carbon film. The analyses were carried out by means of a Hitachi S-3400N SEM with an attached EDX spectrometer, using Oxford Instruments INCA software and the following standards and lines: diopside (Mg$K\alpha$), metal Ni (Ni$K\alpha$), anhydrite (S$K\alpha$), orthoclase (K$K\alpha$), chkalovite (Na$K\alpha$) and TiN (N$K\alpha$). With the used EDX setup (acceleration voltage 20 kV, beam current 1 nA, beam diameter 2 μm), the mineral could withstand until full decomposition for ~30 seconds per analytical point. However, during this time, it loses ~70 relative % of initial ammonium (Supplementary Table S6). Therefore, electron microprobe data can be assessed as semi-quantitative. However, the goals of the analysis were successfully achieved. (1) The presence of nitrogen (N-$K\alpha_2$ line at 392.4 eV) was confirmed (Fig. 1b), consistent with the presence of ammonium determined by both X-ray structural analysis and infrared spectroscopy. The contents of potassium and sodium substituting for ammonium were found insignificant (0.4 wt.% K, 0.1 wt.% Na). (2) No divalent metal cations except for Mg$^{2+}$ and Ni$^{2+}$ were detected. The atomic ratio of Mg to Ni, the only non-volatile constituents, was found almost identical to that determined by X-ray structure refinement (see the main text). The results of semi-quantitative EDX (4 analytical points) are given in Supplementary Table S6.

**Infrared spectroscopy**

FT-IR spectra of the mineral were obtained using a Bruker Hyperion 2000 IR-microscope with a liquid-nitrogen cooled MCT detector attached to a Bruker Vertex 70 spectrometer. The mid-IR spectrum in transmission mode (Fig. 2a) was recorded from a 30 μm crystal which was placed onto and than smeared over the flat polished surface of KBr crystal disc. The 15× reflector objective and 15 μm rectangular field aperture was used. The spectra were recorded over the range of 4000 to 600 cm$^{-1}$ with spectral resolution of 4 cm$^{-1}$. The resultant profile was obtained by averaging of 320 scans. FT-IR spectrum in reflectance mode (Fig. 2b) was obtained from the same mineral film smeared over the KBr disc. The spectrum was recorded using the same instrument setup, but the KBr crystal was placed on a liquid-nitrogen cooled metal table, which allowed to cool the sample and recording the spectrum at 200±20 K. All spectra were processed using Bruker OPUS v. 6.5 software.

**Data availability**

All reported data are available in the article text and in the Supplementary Information. Full crystallographic information can be retrieved from the Crystallographic Information File (CIF), which is available free of charge upon request at the Cambridge Crystallographic Data Centre (CCDC) (www.ccdc.cam.ac.uk/data_request/cif) under deposition number 2374362.

**References**

1. Grewal, D. S., Dasgupta, R. & Marty, B. A very early origin of isotopically distinct nitrogen in inner Solar System protoplanets. *Nat. Astron.* **5**, 356–364 (2021).

2. Pizzarello, S. & Williams, L. B. Ammonia in the early Solar System: an account from carbonaceous meteorites. *Astrophys. J.* **749**, 161–167 (2012).

3. Pizzarello, S. & Bose, M. The path of reduced nitrogen toward early Earth: the cosmic trail and its solar shortcuts. *Astrophys. J.* **814**, 107–114 (2015).

4. Altwegg, K., Balsiger, H. & Fuselier, S. A. Cometary chemistry and the origin of icy solar system bodies: the view after Rosetta. *Ann. Rev. Astron. Astrophys.* **57**, 113–55 (2019).

5. Poch, O. *et al*. Ammonium salts are a reservoir of nitrogen on a cometary nucleus and possibly on some asteroids. *Science* **367**, 6483 (2020).

6. King, T., Clark, R., Calvin, W., Sherman, D. & Brown, R. Evidence for ammonium-bearing minerals on Ceres. *Science* **255**, 1551–1553 (1992).

7. De Sanctis, M. C. *et al*. Ammoniated phyllosilicates with a likely outer Solar System origin on (1) Ceres. *Nature* **528**, 241–244 (2015).

8. Lewis, Z. M., Beth, A., Altwegg, K., Galand, M., Goetz, C., Heritier, K., O'Rourke, L., Rubin, M. & Stephenson, P. Origin and trends in $NH_4^+$ observed in the coma of 67P. *Month. Not. Royal Astron. Soc.* **523**, 6208–6219 (2023)

9. Marion, G., Kargel, J., Catling, D. & Lunine, J. Modeling ammonia–ammonium aqueous chemistries in the Solar System's icy bodies. *Icarus* **220**, 932–946 (2012).

10. Berg, B. L., Cloutis, E. A., Beck, P., Vernazza, P., Bishop, J. L., Driss, T., Reddy, V., Applin, D. & Mann, P. Reflectance spectroscopy (0.35-8 μm) of ammonium bearing minerals and qualitative comparison to Ceres-like asteroid. *Icarus* **265**, 218–237 (2016).

11. Matsumoto, T. *et al*. Influx of nitrogen-rich material from the outer Solar System indicated by iron nitride in Ryugu samples. *Nat. Astron.* **8**, 207–215 (2024).

12. Rubin, A.E. & Ma, C. Meteoritic minerals and their origins. *Geochemistry* **77**, 325–385 (2017).

13. Pilorget, C. *et al*. First compositional analysis of Ryugu samples by MicrOmega hyperspectral microscope. *Nat. Astron.* **6**, 221–225 (2022).

14. Schmitt-Kopplin, P. *et al*. Soluble organic matter Molecular atlas of Ryugu reveals cold hydrothermalism on C-type asteroid parent body. *Nat. Commun.* **14**, 6525 (2023).

15. Jin, S., Hao, M., Guo, Z. *et al*. Evidence of a hydrated mineral enriched in water and ammonium molecules in the Chang'e-5 lunar sample. *Nat. Astron.* **9**, doi: 10.1038/s41550-024-02306-8 (2024).

16. Laize-Générat, L., Soussaintjean, L., Poch, O., Bonal, L., Savarino, J., Caillon, N., Ginot, P., Vella, A., Flandinet, L., Gounelle, M. & Bizzaro, M. Ammonium is a significant reservoir of nitrogen in the Orgueil meteorite. *Europlanet Sci. Congr. 2022*, Granada, Spain, 18–23 Sep 2022, EPSC2022-837, https://doi.org/10.5194/epsc2022-837 (2022).

17. Gounelle, M. & Zolensky, M. E. The Orgueil meteorite: 150 years of history. *Meteor. Planet. Sci.* **49**, 1769–1794 (2014).

18. Court, R.W. & Sephton, M.A. New estimates of the production of volatile gases from ablating carbonaceous micrometeoroids at Earth and Mars during an E-belttype Late Heavy Bombardment. *Geochim. Cosmochim. Acta* **145**, 175–205 (2014).

19. Remusat, L., Derenne, S., Robert, F. & Knicker, H. New pyrolytic and spectroscopic data on Orgueil and Murchison insoluble organic matter: a different origin than soluble? *Geochim. Cosmochim. Acta* **69**, 3919–3932 (2005).

20. Aponte, J. C., Dworkin, J. P. & Elsila, J. E. Indigenous aliphatic amines in the aqueously altered Orgueil meteorite. *Meteor. Planet. Sci.* **50**, 1733–1749 (2015).

21. Ito, M. *et al*. A pristine record of outer Solar System materials from asteroid Ryugu's returned sample. *Nat. Astron.* **6**, 1163–1171 (2022).

22. Yokoyama, T., Nagashima, K. *et al*. Samples returned from the asteroid Ryugu are similar to Ivuna-type carbonaceous meteorites. *Science* **379**, 786 (2023).

23. Lauretta, D. S. *et al*. Asteroid (101955) Bennu in the laboratory: Properties of the sample collected by OSIRIS-Rex. *Meteor. Planet. Sci.* **59** (2024).

24. Grady, M. M. Catalogue of Meteorites, 5th Edn (Natural History Museum, London, 2000).

25. Lodders, K. Solar System abundances and condensation temperatures of the elements. *Astrophys. J.* **591**, 1220–1247 (2003).

26. Tomeoka, K. & Buseck, P. R. Matrix mineralogy of the Orgueil CI carbonaceous chondrite. Geochim. Cosmochim. Acta **52**, 1627–1640 (1988).

27. Fredriksson, K. & Kerridge, J. F. Carbonates and sulfates in CI chondrites: Formation by aqueous activity on the parent body. *Meteoritics* **23**, 35–44 (1988).

28. Daubrée, G.-A. Nouveaux renseignements sur le bolide du 14 mai 1864. *Compt. Rend. Acad. Sci. Paris* **58**, 1065–1072 (1864).

29. Cloëz, S. Analyse chimique de la pierre météorique d'Orgueil. *Compt. Rend. Acad. Sci. Paris* **59**, 37–40 (1864).

30. Urey, H. C. Biological material in meteorites: A review. *Science* **151**, 157–166 (1966).

31. Tutton, A. E. XXIV.—Connection between the atomic weight of contained metals and the magnitude of the angles of crystals of isomorphous series. A study of the potassium, rubidium, and cæsium salts of the monoclinic series of double sulphates $R_2M(SO_4)_2, 6H_2O$. *J. Chem. Soc. Trans.* **63**, 337–423 (1893).

32. Bosi, F., Belardi, G. & Ballirano, P. Structural features in Tutton's salts $K_2[M^{2+}(H_2O)_6](SO_4)_2$, with $M^{2+}$ = Mg, Fe, Co, Ni, Cu, and Zn. *Am. Mineral.* **94**, 74–82 (2009).

33. Diego Gatta, G., Guastella, G., Guastoni, A., Gagliardi, V., Cañadillas-Delgado, L. & Fernandez-Diaz, M. T. A neutron diffraction study of boussingaultite, $(NH_4)_2[Mg(H_2O)_6](SO_4)_2$. *Am. Mineral.* **108**, 354–361 (2023).

34. Culka, A., Jehlička, J. & Němec, I. Raman and infrared spectroscopic study of boussingaultite and nickelboussingaultite. *Spectrochim. Acta A* **73**, 420–423 (2009).

35. Gounelle, M. & Zolensky, M. E. A terrestrial origin for sulfate veins in CI1 chondrites. *Meteor. Planet. Sci.* **36**, 1321–1329 (2001).

36. Airieau, S. A., Farquhar, J., Thiemens, M. H., Leshin, L. A., Bao, H. & Young, E. Planetesimal sulfate and aqueous alteration in CM and CI carbonaceous chondrites. *Geochim. Cosmochim. Acta* **69**, 4166–4171 (2005).

37. Naraoka, H. *et al*. Soluble organic molecules in samples of the carbonaceous asteroid (162173) Ryugu. *Science* **379**, eabn9033 (2023).

38. Rubin, M., Engrand, C., Snodgrass, C., Weissman, P., Altwegg, K., Busemann, H., Morbidelli, A. & Mumma, M. On the Origin and Evolution of the Material in 67P/Churyumov-Gerasimenko. *Space Sci. Rev.* **216**, 102 (2020).

39. Raponi, A. *et al.* Infrared detection of aliphatic organics on a cometary nucleus. *Nat. Astron.* **4**, 500–505 (2020).

40. Altwegg, K. et al. Evidence of ammonium salts in comet 67P as explanation for the nitrogen depletion in cometary comae. *Nat. Astron.* **4**, 533–540 (2020).

41. Chukanov, N. V. IR Spectra of Minerals and Reference Samples Data: Infrared Spectra of Mineral Species. Springer, The Netherlands, 2014.

42. Capaccioni, F. *et al*. The organic-rich surface of comet 67P/Churyumov–Gerasimenko as seen by VIRTIS/Rosetta. *Science* **347**, aaa0628 (2015).

43. Filacchione, G. *et al*. Comet 67P/CG nucleus composition and comparison to other comets. *Space Sci. Rev.* **215**, 19 (2019).

44. Westphal, A. J., Fakra, S. C., Gainsforth, Z., Marcus, M. A., Ogliore, R. C. & Butterworth, A.L. Mixing fraction of inner solar system material in Comet 81P/Wild2. *Astrophys. J.* **694**, 18–28 (2009).

45. Berger, E. L., Zega, T. J., Keller, L. P. & Lauretta, D. S. Evidence for aqueous activity on comet 81P/Wild 2 from sulfide mineral assemblages in Stardust samples and CI chondrites. *Geochim. Cosmochim. Acta* **75**, 3501–3513 (2011).

46. Berger, E. L., Lauretta, D. S., Zega, T. J. & Keller, L. P. Heterogeneous histories of Ni-bearing pyrrhotite and pentlandite grains in the CI chondrites Orgueil and Alais. *Meteor. Planet. Sci.* **51**, 1813–1829 (2016).

47. Yakhontova, L. K., Sidorenko, G. A., Stolyarova, T. I., Plyusnina, I. I. & Ivanova, T. L. Nickel-containing sulfates from the oxidation zone of the Noril'sk deposits. *Zap. Vsesoyuzn. Mineral. Obshchest.* **105**, 710–720 (1976) (in Russian).

48. Chukanov, N. V., Pekov, I. V., Belakovskiy, D. I., Britvin, S. N., Stergiou, V., Voudouris, P. & Magganas, A. Katerinopoulosite, $(NH_4)_2Zn(SO_4)_2·6H_2O$, a new mineral from the Esperanza mine, Lavrion, Greece. *Eur. J. Mineral.* **30**, 821–826 (2018).

49. Bieler, A. et al. Abundant molecular oxygen in the coma of comet 67P/Churyumov-Gerasimenko. *Nature* **526**, 678–681 (2015)

50. Sutton, S., Alexander, C. M. O'D., Bryant, A., Lanzirotti, A., Newville, M. & Cloutis, E. A. The bulk valence state of Fe and the origin of water in chondrites. *Geochim. Cosmochim. Acta* **211**, 115–132 (2017).

51. Macdougall, J. D. & Kerridge, J. F. Cubanite – New sulfide phase in CI meteorites. *Science* **197**, 561–562 (1977).

52. Bullock, E. S., Gounelle, M., Lauretta, D. S., Grady, M. M. & Russell, S. S. Mineralogy and texture of Fe–Ni sulfides in CI1 chondrites: clues to the extent of aqueous alteration on the CI1 parent body. *Geochim. Cosmochim. Acta* **69**, 2687–2700 (2005).

53. Sheldrick, G. M. Crystal structure refinement with SHELXL. *Acta Crystallographica C* **71**, 3–8 (2015).

54. Dolomanov, O.V., Bourhis, L.J., Gildea, R.J., Howard, J.A. & Puschmann, H. OLEX2: a complete structure solution, refinement and analysis program. *J. Appl. Cryst.* **42**, 339–341 (2009).

**Acknowledgements**

The authors are thankful to Resource center of X-ray diffraction studies and Geomodel resource center of St. Petersburg State University for the access to instrumental and computational facilities. This work was financially supported by Russian Science Foundation, grant 24-17-00228.

**Author contributions**

S.N.B. created the concept of the work, selected the mineral, carried out X-ray and IR study, and wrote the manuscript. O.S.V and N.S.V. performed electron microprobe analyses. M.G.K. participated in X-ray diffraction studies. M.A.I. analyzed the data and wrote the manuscript.

**Competing interests**

The authors declare no competing interests.

**Additional information**

**Supplementary information.**

The online version contains supplementary material available at

.

**Correspondence and requests for materials** should be addressed to Sergey Britvin.

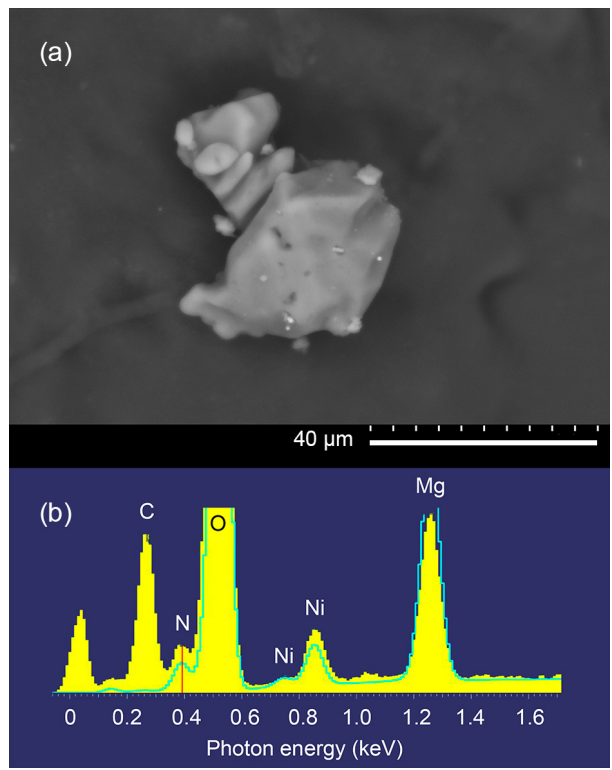

**Fig. 1.** Nickeloan boussingaultite (the $NH_4$-Mg-Ni Tutton salt) handpicked from Orgueil. **a**, General view of the crystal intergrowth in scanning electron microscope (BSE image). **b**, Low-energy region of the EDX spectrum acquired from the same crystal, evident for the presence of nitrogen, magnesium and nickel.

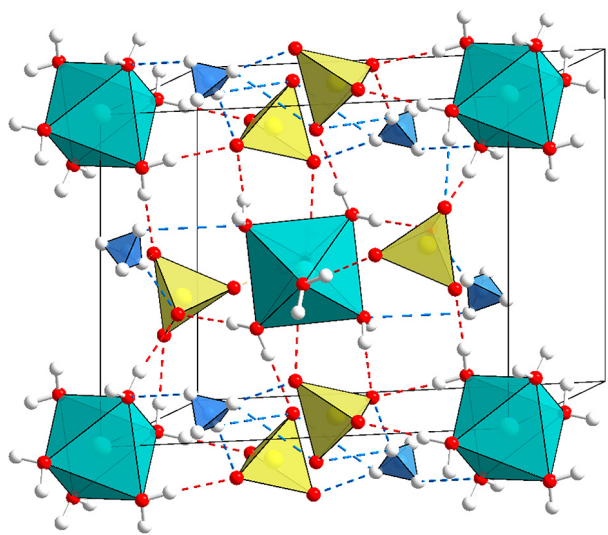

**Fig. 2.** The crystal structure of nickeloan boussingaultite from Orgueil. Isolated octahedra $[(Mg,Ni)(H_2O)_6]^{2+}$ (turquoise), sulfate tetrahedra, $(SO_4)^{2-}$ (yellow) and ammonium tetrahedra, $NH_4^+$ (small blue) cross-linked by a network of hydrogen bonds (dashed lines).

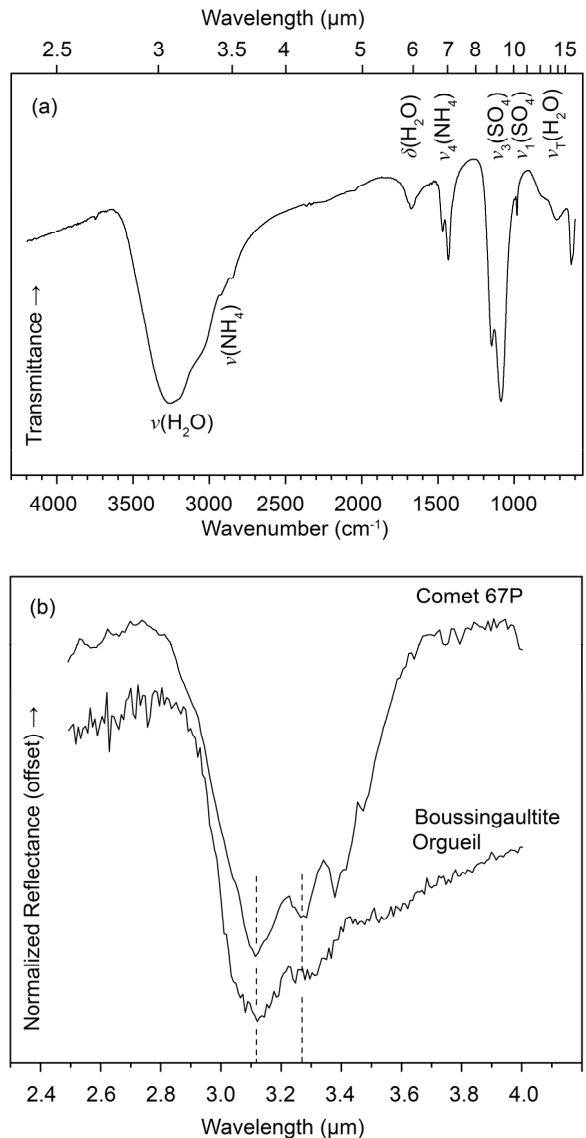

**Fig. 3.** Infrared spectra of nickeloan boussingaultite and reflectance spectrum of comet 67P in the vicinity of a 3.2 μm feature. **a**, Mid-IR spectrum of the mineral recorded in transmission mode at ambient conditions. **b**, The 2.5–4.0 μm region of IR spectra of the mineral (acquired at 200 K) and comet 67P/Churyumov-Gerasimenko. Data for comet 67P were taken from ref. 39.

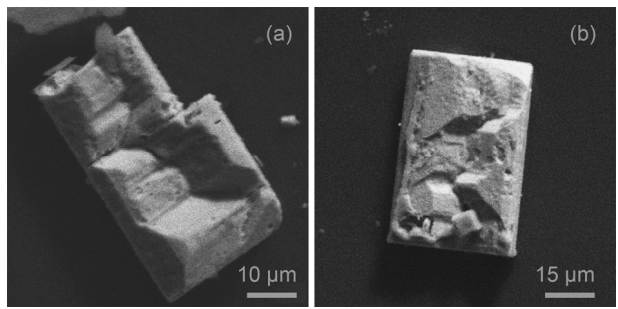

**Fig. 4.** Twinned fresh cubanite crystals extracted from the matrix of Orgueil.

**Table 1**. Band assignments in the infrared transmission spectrum of nickeloan boussingaultite from Orgueil.

| Wavenumber (cm$^{-1}$) [a] | Wavelength (μm) | Assignment [b] |
|---|---|---|
| 3260 s | 3.07 s | $\nu(H_2O)$ |
| 3066 s | 3.26 s | $\nu(H_2O)$ |
| 2925 m sh | 3.42 m sh | $\nu(NH_4)$ |
| 2850 m sh | 3.51 m sh | $\nu(NH_4)$ |
| 1676 m | 5.97 m | $\nu_2(NH_4)+\delta(H_2O)$ |
| 1470 m | 6.8 m | $\nu_4(NH_4)$ |
| 1431 s | 6.99 s | $\nu_4(NH_4)$ |
| 1148 s | 8.71 s | $\nu_3(SO_4)$ |
| 1086 s | 9.21 s | $\nu_3(SO_4)$ |
| 982 m | 10.18 m | $\nu_1(SO_4)$ |
| 816 m sh | 12.25 m sh | $\nu_T(H_2O)$ |
| 722 m | 13.85 m | $\nu_T(H_2O)$ |
| 626 m | 15.97 m | $\nu_4(SO_4)$ |
| 615 m sh | 16.26 m sh | $\nu_4(SO_4)$ |

[a] Intensity and shape abbreviations: s – strong; m – medium; sh – shoulder.
[b] Band assignments according to Ref. 34.

**Table 2**. Ammonium- and hydroxyl-stretching vibrations (the 3.2 μm feature) in the infrared spectra of nickeloan boussingaultite and Comet 67P.

| Orgueil | | Comet 67P | Assignment |
|---|---|---|---|
| Transmittance | Reflectance | Reflectance | |
| 3.07 | 3.13 | 3.11 | $\nu(H_2O)$ |
| 3.26 | 3.29 | 3.26 | $\nu(H_2O)$ |
| 3.42 | 3.46 | 3.38 | $\nu(NH_4)$ |
| 3.51 | 3.53 | 3.48 | $\nu(NH_4)$ |

# Meteoritic Tutton salt, a naturally inspired reservoir of cometary and asteroidal ammonium

Sergey N. Britvin[1,2]*, Oleg S. Vereshchagin[1], Natalia S. Vlasenko[1], Maria G. Krzhizhanovskaya[1] & Marina A. Ivanova[3]

[1]Saint-Petersburg State University, Universitetskaya Nab. 7/9, 199034 St. Petersburg, Russia
[2]Kola Science Center, Russian Academy of Sciences, Fersman Str. 14, 184200 Apatity, Russia
[3]Vernadsky Institute of Geochemistry of the Russian Academy of Sciences, Kosygin St. 19, Moscow 119991, Russia

**Table S1**. Crystal parameters, data collection and structure refinement details for nickeloan boussingaultite from Orgueil.

| **Crystal Data** | |
|---|---|
| Structural formula | $(NH_4)_2[(Mg_{0.65}Ni_{0.35})(H_2O)_6](SO_4)_2$ |
| Crystal size (mm) | $0.03 \times 0.02 \times 0.01$ |
| Crystal system | Monoclinic |
| Space group | $P2_1/c$ (# 14) |
| $a, b, c$ (Å) | 6.2137(3), 12.5213(7), 9.2526(5) |
| $\beta$ (°) | 106.934(6) |
| $V$ (Å$^3$) | 688.67(7) |
| $Z$ | 2 |
| $D_x$ (g cm$^{-3}$) | 1.797 |
| **Data collection and refinement** | |
| Radiation | Mo$K\alpha$ ($\lambda$ = 0.71073 Å) |
| Temperature ($K$) | 293 |
| $2\Theta_{max}$ (°) | truncated at 52 |
| Total reflections collected | 8685 |
| No. of unique reflections | 1360 |
| No. of unique observed, $I > 2\sigma(I)$ | 1175 |
| $h, k, l$ range | $-7\rightarrow7, -15\rightarrow15, -11\rightarrow11$ |
| $F(000)$ | 391 |
| $\mu$ (mm$^{-1}$) | 0.95 |
| No. of refined parameters | 129 |
| $R_{int.}, R_\sigma$ | 0.064, 0.035 |
| $R_1$ [$F \geq 4\sigma(F)$], $wR_2$ | 0.029, 0.072 |
| $S = GoF$ | 1.09 |
| H-atom treatment | All H-atom parameters refined |
| Data completeness | 0.998 |
| Residual density (e Å$^{-3}$) (min, max) | $-0.31, 0.26$ |

**Table S2**. Fractional atomic coordinates and isotropic or equivalent isotropic displacement parameters (Å$^2$) for nickeloan boussingaultite from Orgueil

| Site | x | y | z | $U_{iso}$*/$U_{eq}$ | Occupancy (<1) |
|---|---|---|---|---|---|
| M | 0 | 0 | 0 | 0.0176(2) | Mg 0.650(4) Ni 0.350(4) |
| O1 | 0.2982(2) | −0.06777(14) | −0.0006(2) | 0.0262(4) | |
| O2 | 0.1643(3) | 0.10595(13) | 0.16799(19) | 0.0293(4) | |
| H1A | 0.327(5) | −0.130(3) | 0.026(3) | 0.043(8)* | |
| H1B | 0.336(5) | −0.064(2) | −0.080(4) | 0.050(9)* | |
| H2A | 0.282(5) | 0.094(2) | 0.205(3) | 0.035(8)* | |
| H2B | 0.110(5) | 0.122(2) | 0.236(3) | 0.042(8)* | |
| O3 | 0.0324(3) | 0.10941(14) | −0.16058(19) | 0.0278(4) | |
| H3A | 0.010(4) | 0.172(2) | −0.144(3) | 0.033(7)* | |
| H3B | −0.041(5) | 0.099(2) | −0.249(4) | 0.044(8)* | |
| S1 | 0.25971(8) | 0.86179(4) | 0.59393(5) | 0.02196(18) | |
| O4 | 0.3750(2) | 0.93321(12) | 0.71959(16) | 0.0282(4) | |
| O5 | 0.0487(2) | 0.82216(13) | 0.61743(18) | 0.0318(4) | |
| O6 | 0.4108(3) | 0.77117(13) | 0.59175(19) | 0.0360(4) | |
| O7 | 0.2141(3) | 0.92041(15) | 0.45261(18) | 0.0434(5) | |
| N1 | 0.3567(4) | 0.3499(2) | 0.1316(3) | 0.0267(4) | |
| H1C | 0.238(7) | 0.342(3) | 0.080(4) | 0.064(12)* | |
| H1D | 0.361(5) | 0.410(3) | 0.159(3) | 0.046(9)* | |
| H1E | 0.441(6) | 0.342(3) | 0.087(4) | 0.053(10)* | |
| H1F | 0.387(5) | 0.313(3) | 0.197(4) | 0.053(10)* | |

**Table S3**. Anisotropic displacement parameters (Å$^2$) for nickeloan boussingaultite from Orgueil

| Site | $U^{11}$ | $U^{22}$ | $U^{33}$ | $U^{12}$ | $U^{13}$ | $U^{23}$ |
|---|---|---|---|---|---|---|
| *M* | 0.0175 (3) | 0.0190 (4) | 0.0156 (3) | 0.0007 (2) | 0.0039 (2) | –0.0008 (2) |
| O1 | 0.0258 (8) | 0.0260 (9) | 0.0296 (9) | 0.0049 (6) | 0.0123 (7) | 0.0032 (7) |
| O2 | 0.0233 (9) | 0.0363 (10) | 0.0260 (9) | –0.0003 (7) | 0.0034 (8) | –0.0085 (7) |
| O3 | 0.0372 (9) | 0.0245 (9) | 0.0215 (9) | 0.0011 (7) | 0.0085 (7) | 0.0017 (7) |
| S1 | 0.0225 (3) | 0.0233 (3) | 0.0189 (3) | –0.0021 (2) | 0.0043 (2) | –0.00308 (19) |
| O4 | 0.0294 (8) | 0.0300 (9) | 0.0236 (8) | –0.0044 (6) | 0.0051 (6) | –0.0074 (6) |
| O5 | 0.0245 (8) | 0.0305 (8) | 0.0424 (9) | –0.0036 (6) | 0.0127 (7) | –0.0041 (7) |
| O6 | 0.0300 (8) | 0.0285 (9) | 0.0501 (10) | 0.0021 (7) | 0.0125 (7) | –0.0112 (7) |
| O7 | 0.0523 (11) | 0.0504 (11) | 0.0226 (8) | –0.0060 (9) | 0.0035 (7) | 0.0049 (8) |
| N1 | 0.0257 (11) | 0.0296 (13) | 0.0268 (11) | 0.0013 (9) | 0.0110 (10) | 0.0004 (10) |

**Table S4**. Interatomic bond lengths (Å) in nickeloan boussingaultite from Orgueil

| Bond | Length | Bond | Length |
|---|---|---|---|
| *M*–O1 | 2.0396(15) | S1–O4 | 1.4764(15) |
| *M*–O2 | 2.0721(16) | S1–O5 | 1.4758(16) |
| *M*–O3 | 2.0734(16) | S1–O6 | 1.4765(16) |
| O1–H1A | 0.82(3) | S1–O7 | 1.4536(17) |
| O1–H1B | 0.84(3) | N1–H1C | 0.76(4) |
| O2–H2A | 0.73(3) | N1–H1D | 0.79(3) |
| O2–H2B | 0.82(3) | N1–H1E | 0.76(4) |
| O3–H3A | 0.82(3) | N1–H1F | 0.74(4) |
| O3–H3B | 0.82(3) | | |

**Table S5**. Hydrogen-bond geometry (Å, °) for nickeloan boussingaultite from Orgueil [a]

| *D*–H⋯*A* | *D*–H | H⋯*A* | *D*⋯*A* | *D*–H⋯*A* |
|---|---|---|---|---|
| O1–H1*A*⋯O6[i] | 0.82(3) | 1.89(3) | 2.712(2) | 177(3) |
| O1–H1*B*⋯O4[ii] | 0.84(3) | 1.94(4) | 2.766(2) | 170(3) |
| O2–H2*A*⋯O4[iii] | 0.73(3) | 2.07(3) | 2.790(2) | 172(3) |
| O2–H2*B*⋯O5[iv] | 0.82(3) | 2.02(3) | 2.835(2) | 171(3) |
| O3–H3*A*⋯O5[v] | 0.82(3) | 1.94(3) | 2.762(2) | 176(2) |
| O3–H3*B*⋯O7[vi] | 0.82(3) | 1.89(3) | 2.716(2) | 178(3) |
| N1–H1*C*⋯O5[v] | 0.76(4) | 2.16(4) | 2.897(3) | 163(4) |
| N1–H1*D*⋯O4[vii] | 0.79(3) | 2.03(4) | 2.828(3) | 177(3) |
| N1–H1*E*⋯O6[viii] | 0.76(4) | 2.29(4) | 3.011(3) | 157(3) |
| N1–H1*F*⋯O6[iii] | 0.74(4) | 2.25(4) | 2.964(3) | 161(3) |

[a] Symmetry codes: (i) *x*, -*y*+1/2, *z*-1/2; (ii) *x*, *y*-1, *z*-1; (iii) -*x*+1, -*y*+1, -*z*+1; (iv) -*x*, -*y*+1, -*z*+1; (v) -*x*, *y*-1/2, -*z*+1/2; (vi) -*x*, -*y*+1, -*z*; (vii) *x*, -*y*+3/2, *z*-1/2; (viii) -*x*+1, *y*-1/2, -*z*+1/2.

**Table S6.** Assessed chemical composition of nickeloan boussingaultite from Orgueil (electron microprobe, wt. %)

| Point | (NH$_4$)$_2$O | K$_2$O | Na$_2$O | MgO | NiO | SO$_3$ | H$_2$O |
|---|---|---|---|---|---|---|---|
| 1 | 2.71 | 0.41 | 0.11 | 4.01 | 8.17 | 33.86 | |
| 2 | 4.78 | 0.42 | 0.00 | 6.70 | 7.61 | 39.18 | |
| 3 | 5.60 | 0.34 | 0.15 | 9.54 | 6.81 | 42.53 | |
| 4 | 2.86 | 0.41 | 0.00 | 6.67 | 7.28 | 38.88 | |
| Average | 3.99 | 0.40 | 0.07 | 6.73 | 7.47 | 38.61 | 42.74 [a] |
| Theoretical | 13.65 | 0.38 | 0.08 | 6.79 | 7.39 | 42.81 | 28.90 |
| Formula amounts based on (Mg+Ni)=1 (atoms per formula unit) | | | | | | | |
| Average | 0.57 | 0.03 | 0.01 | 0.63 | 0.37 | 1.81 | 8.88 |
| Theoretical | 1.96 | 0.03 | 0.01 | 0.63 | 0.37 | 2.00 | 6.00 |

[a] Water by difference.